\documentclass[preprint]{aastex}
\usepackage{psfig}
\newcommand{\lumimoo}{$10^{-11}~{\rm ergs}~{\rm s}^{-1}$~cm$^{-2}$}

\newcommand{\MTrelation}{$M_{\rm tot}-T_{X}$~}

\shorttitle{Mass-Temperature Relation}
\shortauthors{Xu et al.}

\begin{document}

\title{\vspace{-1cm} The Mass-Temperature Relation of 22 Nearby Clusters}

\author{
{\sc Haiguang} {\sc Xu}$^{1}$, {\sc Guangxue} {\sc Jin}$^{1}$ and
{\sc Xiang-Ping} {\sc Wu}$^{2}$ \\
{\small\it $^{1}$ Department of Applied Physics, Shanghai Jiao Tong University,} \\{\small\it Huashan Road 1954, Shanghai 200030, PRC} \\
{\small\it $^{2}$Beijing Astronomical Observatory and National Astronomical
Observatories, Chinese Academy of Sciences, 
20A Datun Road, Beijing 100012, PRC}\\}

\begin{abstract}
We present a new investigation of the mass-temperature 
(\MTrelation) relation of 22 nearby clusters based on 
the analysis of their ROSAT X-ray surface brightness 
profiles ($S_{X}(r)$) and their ASCA emission 
weighted temperatures. Two methods of the cluster mass 
estimations are employed and their results are compared:  
(1) the conventional $\beta$ model for gas distribution
along with the isothermal and hydrostatic equilibrium 
assumptions, and (2) the NFW profile for dark matter 
distribution whose characteristic density and length 
are determined by the observed $S_{X}(r)$. These two 
models yield essentially the same goodness of fits for
$S_{X}(r)$ and the similar \MTrelation relations, with 
the latter demonstrating a significant departure from 
the simple gravitational scaling of $M_{tot}\propto T_X^{3/2}$.
It is also shown that the best-fit \MTrelation relations 
could be reconciled with the theoretical expectation if 
the low-temperature clusters ($T_X<3.5$ keV) are excluded 
from the list, which lends support to the scenario that the 
intra-cluster medium is preheated in the early phase of 
cluster formation. Together with the entropy-temperature 
distribution, the existence of a similarity break at 
$T_X=3-4$ keV in the dynamical scaling relations for 
galaxy clusters has been confirmed. 
\end{abstract}


\section{Introduction}

In the past ten years, the ROSAT and ASCA observations
of galaxy clusters have revealed that there exist 
tight correlations between the total gravitating mass of 
clusters ($M_{tot}$) and the X-ray 
luminosity ($L_{X}$) and the temperature ($T_{X}$) of the intra-cluster 
medium (ICM) (e.g. David et al., 1993; White, Jones \& Forman 1997;
Markevitch 1998; Wu, Xue \& Fang, 1999; Horner, 
Mushotzky \& Scharf 1999; etc.).  Presence of 
such relations has been also motivated by both the
theoretical studies and the $N$-body numerical simulations. 
A close comparison of the theoretical/simulated 
results and the observed data can therefore provide 
valuable constraints on the prevailing cosmological models and 
even on the nature of dark matter which 
dominates the mass distribution and the dynamical evolution of clusters.
However, in addition to its dependence on the hydrostatic equilibrium
hypothesis, an accurate determination of $M_{tot}$ often suffers 
from instrumental limitations, especially in those faint and distant 
clusters. This compares with the ICM temperature measurements which
can be relatively easily performed 
through the X-ray spectroscopy in most clusters. Consequently, 
a well-established  \MTrelation relation can instead be used as a powerful 
cluster mass estimator.  In particular, 
based on the \MTrelation curve, one is able to  
link the observable quantities with the 
theoretical Press-Schechter mass function, which appears to be 
crucial for the determinations of the cosmological parameters using 
the cluster abundances at different redshifts.

Within the framework of pure gravitational infall (e.g., Lilje 1992), 
the cluster mass and the gas temperature should scale simply as
$M_{tot} \propto T_{X}^{\alpha}$, 
where the index $\alpha~(= 3/2)$ 
is almost unrelated to the different settings of the 
cosmological parameters. This self-similarity has been justified  
by cosmological $N$-body simulations (Evrard et al. 1996; Bryan 
\& Norman 1998; Eke et al. 1998), and has also been tested by numerous
observations. For example, Hjorth et al. 
(1998) have inferred the cluster mass from the weak and strong 
lensing for 8 clusters and confirmed such a tight mass-temperature 
relation. Neumann et al. (1999) analyzed the ROSAT PSPC and HRI 
data of 26 clusters and estimated the cluster mass by fitting the 
X-ray brightness profiles with the isothermal $\beta$-model, 
coupled with the hydrostatic equilibrium assumption. 
For a sub-sample of 15 targets whose temperature are available, 
they achieved a result consistent with 
$M_{tot} \propto T_{X}^{3/2}(1+z)^{-3/2}$.

However, it should be noticed that in the above observational 
analyses, only high-temperature ($kT>4-5$ keV) clusters are involved. 
As has been realized in many recent studies (e.g. Mohr, Mathiesen
\& Evrard 1999; Ponman, Cannon \& Navarro 1999),  
the influence of energy feed-back into ICM 
in the low-temperature clusters/groups can become more significant than 
that in the high-temperature systems. Consequently, the self-similarity 
may break at the low-temperature end. Indeed, this has been confirmed
in Nevalainen et al. (2000) using a small sample of 9 clusters 
(6 clusters at $4$ keV and 3 groups at $\simeq 1$ keV).  From 
the ASCA and the ROSAT temperature profiles, 
the authors have found that for the whole sample  
$M_{tot}$ scales as $T_{X}^{1.79 \pm 0.14}$,
which is apparently steeper than the theoretical prediction, 
whereas $M_{tot} \propto T_{X}^{3/2}$ still holds if 
the low-temperature groups are excluded.

In the scenario of preheated ICM, such a similarity break can
develop naturally in the low-temperature clusters/groups.
A robust constraint on the break temperature ($T_{\rm break}$)
 can be obtained in principle
by the observations of the $L_{X}$-$T_{X}$ or $M_{vir}$-$T_{X}$ 
relations, which will in turn yield valuable information about 
the heating processes in the early phase of cluster formation.
In the recent analysis of Nevalainen et al. (2000), 
only a lower limit of 1 keV could be set to $T_{break}$ 
due to the temperature gap between $1-4.5$ keV in their sample.
On the other hand, Ponman et al. (1999) evaluated the 
entropy of ICM in different systems and argued that $T_{break}$ 
could be as high as $\sim 3$ keV. Yet, this seems to be 
inconsistent with the finding of Horner et al. (1999) that 
the similarity of \MTrelation may continue until $T\sim2$ keV.

In this paper, we report our study of the 
\MTrelation relation using a combined sample of 22 galaxy clusters
observed with ROSAT and ASCA.  
The radial X-ray surface brightness profiles are derived and 
fitted with different models. Based on the best-fit parameters,
the total gravitational masses scaled to the virial radius and 
the ICM entropies at the cluster central regions are calculated 
under the assumption of isothermality. Throughout this paper,
we employ the MEKAL model (Mewe, Kaastra \& Liedahl 1995) 
to calculate the thermal emission 
from the plasma with a mean metallic abundance of 0.4 solar, 
and set the Hubble constant to $50 h_{50}$ km s$^{-1}$ Mpc$^{-1}$.

\section{The Sample and Data Reduction}

Our sample (Table \ref{tbl:targets}) consists of 22 galaxy 
clusters, whose redshifts range from $0.01$ to $0.09$. At 
this distance range, the effect of cosmic evolution can be 
ignored. ROSAT PSPC data archived at the Institute of High 
Energy Physics (Beijing) are analyzed. In general, the 
observations are chosen in such a way that targets 
are located near the center of FOV, which makes it easier to 
study their diffuse X-ray emission as a whole. All the targets 
are sufficiently bright with their 2--10 keV flux being at least 
several \lumimoo. Thus, with sufficient exposure times, high
signal-to-noise ratio can be achieved. We selected targets
that all demonstrate relatively good circular symmetry in 
their X-ray morphology, and contain no distinct substructures. Such 
systems may have been undisturbed for rather a long time, for
which the assumption of spherical symmetry is more reasonable.

We have searched in the literature for the emission weighted ASCA
temperatures for our targets. Essentially, we choose the temperatures
obtained when the central cluster regions are excluded. This is to
avoid the complication due to the additional cool emission 
component (e.g. Buote 2000). 
When the error bars of the temperatures are asymmetric, 
we adopt the values of the larger sides. Among the 
targets, A262 has the lowest temperature at $kT=2.15 \pm 0.06$ keV, 
and A2029 is the hottest one at about 10 keV. It appears that the sample 
almost covers the whole temperature range of clusters. 
The galaxy groups with even lower temperatures at 
$\simeq 1$ keV are not included in this work.

\section{The Models}

We have extracted the radial X-ray surface brightness profiles 
for the targets by directly using their 0.5--2 keV count 
rates (Figure \ref{fig:f1}).
Two approaches are adopted to analytically approximate 
the observed $S_{X}(r)$: 
the $\chi^2$ fit via the conventional $\beta$ model and the $\chi^2$ fit
via the theoretically expected $S_{X}(r)$ tracing
a priori given dark matter density profile. For the latter, we use 
the cusped NFW profile and the empirical Burkert profile with a 
definite core radius.

\subsection{$\beta$ Model}

If both galaxies and ICM are assumed to be isothermal and  
trace the common gravitational potential in
a cluster, the ICM density $n_{g}$ can be represented by the 
so-called $\beta$ model when the galaxy 
distribution is assumed to follow the King profile 
(Cavaliere \& Fusco-Femiano 1976): 
\begin{equation}
\label{eq:betagas}
n_{g}(R)=n_{g0}
        \left[
        1+(R/R_{c})^{2}
        \right]^{-3\beta/2}.
\end{equation}
This density profile can also be obtained by inversion of the
observed $S_{x}(r)$ in form of 
\begin{equation}
\label{eq:betasurface}
S_{X}(r) = S_0\left[1+\left(\frac{r}{R_c}\right)^2\right]^{-3\beta+\frac{1}{2}},
\end{equation}
where $R_{c}$ is the core radius and $\beta$ is the slope  
parameter. Here, we use $R$ to represent the three-dimensional 
radius and $r$ for the two-dimensional radius, both measured from 
the cluster center. The total cluster mass within radius $R$ can be
derived from the hydrostatic equilibrium assumption as
\begin{equation}
\label{eq:betamass}
M_{\rm tot}(R) = -\frac{kT_{X}R}{\mu m_{\rm p} G}\left\{  \frac{{\rm
dln}n_{g}(R)}{{\rm dln}{R}}+ \frac{{\rm dln}T_{X}(R)}{{\rm dln}{R}} \right\}.
\end{equation}

\subsection{NFW profile}

The second model we apply is the NFW profile, suggested by 
the high resolution $N-$body simulations (Navarro, Frenk, 
\& White 1995): 
\begin{equation}
\label{eq:nfw}
\rho (R) = \frac{\delta_{c}\rho_{crit}} {(R/R_{s})(1+R/R_{s})^{2}},
\end{equation}
where $\rho$ is the halo density, 
$R_{s}$ is the scale radius, $\rho_{crit}$ 
is the critical density of the universe, and $\delta_{c}$ is the 
characteristic density contract, respectively. In 1998,
Makino, Sasaki \& Suto derived an analytic form of  
gas distribution from Eq(\ref{eq:nfw}) by assuming (a) spherical 
symmetry, (b) hydrostatic equilibrium, and (c) that the ICM is 
an ideal gas:
\begin{equation}
\label{eq:NFWgas}
\rho_{g}(R)=\rho_{g0}{\rm exp}(-27b/2)(1+R/R_{s})^{27b/(2R/R_{s})},
\end{equation}
where $b$ is a dimensionless constant. With Eq(\ref{eq:NFWgas}), 
we can straightforwardly calculate the ICM X-ray emission  
and predict the corresponding X-ray surface brightness profile $S_X(r)$.
Fitting this theoretically expected  $S_X(r)$ to the X-ray observed
one, we will be able to check the goodness of fit by the NFW profile and 
simultaneously fix the free parameters $R_{s}$ and $\delta_{c}$.
Finally, we obtain the total cluster mass through
\begin{equation}
M_{\rm tot}(R) = 4\pi \delta_{c} \rho_{\rm crit} R_s^3 
                \left[\ln\left(1+\frac{R}{R_s}\right)-
                \frac{R}{R+R_s}\right].
\end{equation}

\subsection{Burkert profile}

Many recent observations, mainly based on the measurements of the rotation 
curves of dwarf galaxies, have shown that galactic 
halos may have flatter cores rather than the cusped density
profiles such as the NFW profile. 
Indeed, such flat cores may develop if the dark 
matter particles are weakly self-interacting 
(e.g., Spergel \& Steinhardt 1999; Burkert 2000). 
An empirical density form for such halos is proposed by Burkert (1995):
\begin{equation}
\label{eq:burkert}
\rho(R)=\frac{\rho_{0}R_{0}^{3}}{(R+R_{0})(R^{2}+R_{0}^{2})},
\end{equation}
in which $\rho_{0}$ and $R_{0}$ are the central density 
and core radius, respectively. Although Eq(\ref{eq:burkert}) 
was originally suggested for dwarf galaxies, it should
also work for clusters, if the dark halos are self-similar.
Indeed,  the recent simulations of Yoshida et al. (2000) have shown that
Eq(\ref{eq:burkert}) is applicable to clusters,
provided that the halo particles are indeed self-interacting.

Following Makino et al. (1998), we derive the radial gas distribution as
\begin{equation}
\label{eq:burkertgas}
n_{g}(R) = n_{g0} {\rm exp}
         \left\{
            k_{0} {\rm ln} 
              \left[
                \frac  
                {(R+R_{0})^{\frac{2}{R}+\frac{2}{R_{0}}} (R^{2}+R_{0}^{2})^{\frac{1}{R}-\frac{1}{R_{0}}}}
                {R_{0}^{4/R}}
              \right]
            - 2k_{0} {\rm arctg}\left(\frac{R}{R_{0}}\right) 
              \left(\frac{1}{R}+\frac{1}{R_{0}}\right)       
         \right\},
\end{equation}
where $k_{0}$ is a dimensionless constant. Then, in a similar way to
the NFW profile, we can predict the resulting X-ray surface brightness 
profile as a result of thermal emission and compare it with X-ray
observations. This would allow us to check the goodness of fit and work out
the free parameters in the Burkert profile and hence, the total cluster
mass.

\section{Results}

\subsection{Goodness of fittings}

We find that the goodness of fit of $S_{X}(r)$ is almost 
indistinguishable between the $\beta$ model and the NFW 
model (Table \ref{tbl:fits}), but is apparently worse for 
the Burkert model for which $\chi_{r}^{2}$ are larger 
by a factor of 2-3. 
We demonstrate one example (A2199) in Fig. \ref{fig:f2}
by plotting the residuals of the fittings for the Burkert profile. 
It is apparent that the model provides a significant overestimate of 
$S_{X}(r)$  at about $2'$ and an underestimate at the adjacent regions. 
The situation is quite common in our fittings, implying that the Burkert 
model tends to give a core which is too flat. Consequently, 
the failure of the Burkert profile in the fittings of
$S_{X}(r)$ essentially excludes this empirical model in 
our mass estimates of clusters below.

\subsection{The Break of Mass- and Entropy-Temperature Relations}

Based on the best-fits with the $\beta$ model and the NFW 
model, we calculate the total cluster mass within the virial 
radius $r_{200}$, which is defined as the radius where 
the mean cluster density is $200$ times of the critical density of 
the universe. As can be seen in 
Table \ref{tbl:fits}, the values of $r_{200}$ are typically in the range 
1.5--3.5 Mpc which are apparently larger than the observable radii of the
clusters. Here we estimate $r_{200}$ for each cluster using 
the extrapolated mass profile.
The uncertainties in mass are calculated by taking into account
the corresponding measurement errors in both the temperature and 
the surface brightness profile. It is found that the resulted mass 
errors are always dominated by the errors of the surface
brightness profile. This last point is crucial because the estimates
of virial radius and total cluster mass depend on gas temperature 
through equation (3). An apparent correlation between $M_{\rm tot}$ and
$T_X$ may arise if the errors of the gas temperature are relatively large.
The mean error level for our cluster sample is about 20\%, 
which is consistent with the predictions in the numerical studies. 
The resulting mass and temperature 
relation is plotted in Figure \ref{fig:f3}.

For the entire sample, we find that the best-fit mass-temperature 
relations are
\begin{eqnarray}
M_{\rm tot} = 5.75\pm0.03\times 10^{13}h_{50}^{-1} 
 \left(\frac{T_X}{\rm keV}\right)^{1.60\pm0.04}~M_{\rm \odot}, & 
 {\rm (\beta)};\\
M_{\rm tot} = 3.43\pm\times 10^{13}h_{50}^{-1} 
 \left(\frac{T_X}{\rm keV}\right)^{1.81\pm0.04}~M_{\rm \odot}, & {
 \rm (NFW)}.
\end{eqnarray} 
Both of the relations are steeper than the theoretical prediction
of $M\propto T^{3/2}$. Since 
such steepness has been suggested as a result of the pre-heating 
of ICM, which could be more remarkable in the low-temperature 
systems, we have excluded the low-temperature clusters one by one
to examine beyond which break temperature the similarity is 
restored. We find that when those clusters at $kT<3.5$ keV 
are excluded, the \MTrelation scaling become self-similar
again with $\alpha=1.54^{+0.06}_{-0.05}$
and $1.59\pm0.04$
for $\beta$ and NFW model, respectively. This result agrees 
well with the one found by Ponmann et al. (1998), who discovered 
that in ICM at $<3$ keV the gas entropy is significantly  
higher than achievable through gravitational collapse alone.
Recently, Tozzi \& Normal (2000) used a generalized model
to simulate the X-ray properties of the clusters. Their
\MTrelation in the low density, cold dark matter
universe with an initial entropy 
$K=0.4\pm0.2\times10^{34}$ erg cm$^{-2}$ g$^{-5/3}$ 
is again in good accordance with our results, showing that 
the emission weighted temperature of the $kT<2$ keV clusters
are about 25\% higher than what is predicted by the self-similar scaling.

In order to confirm our result of $T_{\rm break}$ obtained via
the \MTrelation,  we calculate the ICM 
entropy at $0.1r_{200}$ and illustrate
the entropy-temperature distribution in Figure 
\ref{fig:f4}, where we use the entropy variable 
$S=T_{X}/n_{g}^{2/3}$. 
It appears that there is a clear tendency
of the entropy increase for clusters at $kT_{X}>4$ keV,  
suggesting again that $T_{\rm break}$ should be between 
3 and 4 keV.

\section{Summaries}

We have analyzed the archival data of ROSAT PSPC pointing 
observations of 22 galaxy clusters with gas temperatures 
ranging from 2 to 10 keV. The radial surface brightness 
profiles are extracted from the 0.5--2 keV count rates, 
and are fitted with the $\beta$ model, the NFW model, 
and the Burkert model, respectively.

We have found that the goodness of fit is indistinguishable 
between the $\beta$ model and the NFW model, but is significantly 
worse for the Burkert model. For the latter,  
the simple reason is that the
Burkert model tends to present too flat a core. If 
the Burkert model can be used to describe the gravitational 
potential wells in both the galactic halos and cluster halos
within the framework of the self-interacting scenario, we 
may conclude that the cross-sections of 
dark matter particles may vary with halo sizes.
Yoshida et al. (2000) have recently arrived at  
the similar conclusion by analyzing their $N$-body simulations.

Based on the best-fit parameters of the $\beta$ model and the 
NFW model, we have calculated the cluster mass within the virial 
radius ($r_{200}$). An examination of the \MTrelation shows that
the \MTrelation is steeper than the self-similarity prediction
unless the clusters at $<3.5$ keV are excluded. A 
further study of the ICM entropy as a function of the temperature 
shows that an entropy floor is likely to present in the clusters
cooler than $4$ keV. Therefore, we conclude that 
the break temperature of the self-similarity is at about 3--4 keV. 
We have noticed that based on a more complicated sample
with spatially resolved  temperature profiles, 
Finoguenov, Reiprich \& B\"{o}hringer (2000) 
have arrived at a similar conclusion.

Finally, our \MTrelation may be moderately affected if the intracluster gas
deviates from the simple isothermality. Indeed, 
previous studies have arrived at conflicting results regarding
the radial temperature gradients in clusters.
For examples, Irwin \& Bregman (2000) and Tamura et al. (2001) 
claimed an essentially flat temperature profile, while 
Markevitch et al. (1998) and Kaastra et al. (2000) reported a noticeable
decline of gas temperature at large radii.  However, 
according to the recent analysis of Finoguenov, Reiprich \& B\"{o}hringer 
(2000), it is unlikely that the \MTrelation can be significantly altered  
even if the temperature gradient is taken into account.

\acknowledgements


This research has made use of ROSAT archival data obtained from the 
Institute of High Energy Physics in Beijing, and is supported by the 
National Science Foundation of China, under Grant No. 19803003 and
19725311, and the Ministry of Science and Technology of China, under 
Grant NKBRSF G19990754.


\newpage 

{\bf Figure Captions}

\figcaption[./f1.ps]{Vignetting-corrected and background-subtracted
   surface brightness profiles of our targets in 0.5--2 keV. \label{fig:f1.ps}}

\figcaption[./f2.ps]{Residuals of the Burkert fitting of the 0.5--2 keV 
   radial surface brightness of A2199. \label{fig:f2.ps}}

\figcaption[./f3a.ps, ./f3b.ps]{The mass-temperature relations
  by the $\beta$ model and NFW model. \label{fig:f3.ps}}

\figcaption[./f4.ps]{Gas entropy calculated at $0.1r_{200}$ as a function
of the gas temperature. Diamonds: our result; Full line: the theoretical
prediction by self-similarity from Ponman, Cannon, Navarro (1999)
   \label{fig:f4.ps}}

\clearpage

{\small
\begin{table}[b]
\caption{Target Properties and the ROSAT Observations}
\label{tbl:targets}
\begin{center}
\begin{scriptsize}
\begin{tabular}{llcccclc}
\hline \hline
Target         &sequence &\multicolumn{2}{c}{Pointing Position in
J2000}&Offset&Exposure & Redshift & Temperature$^{*}$\\
               &
&\multicolumn{1}{c}{RA}&\multicolumn{1}{c}{DEC}&(arcmin)&(ks)  & & (keV) \\
\hline
A85   &rp800250n00&00$^h$41$^m$50$^s$&-09$^d$18$^m$00$^s$& 4.2&10.2& 0.0518 &
$6.31\pm0.25^{1}$\\
A119  &rp800251n00&00~56~17&-01~15~00& 1.3&15.2& 0.0440 & $5.59\pm0.27^{1}$\\
A262  &rp800254n00&01~52~50&~36~09~00& 0.2& 8.7& 0.0161 & $2.15\pm0.60^{1}$\\
A401  &rp800235n00&02~58~58&~13~34~48& 0.4& 7.5& 0.0748 & $8.00\pm0.40^{2}$\\
A478  &wp800193n00&04~13~26&~10~28~12& 0.4&22.0& 0.09   & $6.90\pm0.35^{1}$\\
A496  &rp800024n00&04~33~38&-13~15~36& 0.9& 8.8& 0.0320 & $4.13\pm0.08^{1}$\\
A644  &wp800379n00&08~17~24&-07~30~36& 0.5&10.2& 0.0704 & $7.90\pm0.80^{2}$\\
A1060 &rp800200n00&10~36~43&-27~31~48& 1.7&15.8& 0.0114 & $3.24\pm0.06^{1}$\\
A1651 &wp800353n00&12~59~22&-04~12~00& 0.8& 7.4& 0.0825 & $6.10\pm0.40^{2}$\\
A1795 &rp800105n00&13~48~55&~26~35~24& 0.6&36.3& 0.0616 & $5.88\pm0.14^{1}$\\
A2029 &rp800249n00&15~10~55&~05~45~00& 0.4&12.5& 0.0767 & $9.10\pm1.00^{2}$\\
A2052 &rp800275n00&15~16~43&~07~01~12& 0.3& 6.2& 0.0348 & $2.80\pm0.20^{3}$\\
A2063 &rp800184a01&15~23~05&~08~36~36& 0.1& 9.8& 0.0337 & $3.68\pm0.11^{1}$\\
A2199 &rp800644n00&16~28~38&~39~33~00& 0.1&40.0& 0.0303 & $4.10\pm0.08^{1}$\\
A2319 &rp800073a01&19~21~12&~43~57~36& 1.3& 3.2& 0.0564 & $8.90\pm0.34^{1}$\\
A2597 &rp800112n00&23~25~22&-12~07~12& 1.1& 7.2& 0.0852 & $4.40\pm0.70^{2}$\\
A3266 &rp800552n00&04~31~10&-61~28~48& 2.1&13.5& 0.0594 & $8.00\pm0.50^{2}$\\
A3558 &rp800076n00&13~27~55&-31~29~24& 1.5&29.5& 0.0478 & $5.12\pm0.20^{1}$\\
A3562 &rp800237n00&13~33~38&-31~40~12& 1.6&20.2& 0.0499 & $3.80\pm0.09^{1}$\\
A3571 &rp800287n00&13~48~22&-32~56~24&12.0& 6.1& 0.039  & $6.73\pm0.17^{1}$\\
A4059 &wp800175n00&23~57~00&-34~45~36& 6.7& 5.4& 0.0478 & $3.97\pm0.12^{1}$\\
MKW3S &rp800128n00&15~21~53&~07~42~36& 0.6&10.0& 0.045  & $3.68\pm0.09^{1}$\\
\hline
\end{tabular}
\end{scriptsize}
\end{center}
\begin{description}
  \begin{footnotesize}
  \setlength{\itemsep}{-1mm}
     \item[$*$] The temperatures are quoted from [1] Fukazawa (1997), [2]
  Nevalainen et al. (2000), and [3] Hradecky et al. (2000).
  \end{footnotesize}
  \end{description}
\end{table}
}

\clearpage

{\small
\begin{table}[b]
\caption{Best Fits of the Radial Surface Brightness}
\label{tbl:fits}
\begin{center}
\begin{scriptsize}
\begin{tabular}{lcccccccc}
\hline \hline
Target &\multicolumn{4}{c}{$\beta$-model}&\multicolumn{4}{c}{NFW model}\\
       &$R_{c}$ (kpc)&$\beta$&$\chi^{2}_{r}$&$R_{200}$ (Mpc)&$R_s$&$27b/2$&$\chi^{2}_{r}$&$R_{200}$ (Mpc)\\ 
\hline
A85   &$95.1\pm14.7 $&$0.54\pm0.01$&6.1 &$2.8\pm0.5$&$320.0\pm6.9 $&$7.66\pm0.11$&3.2&$2.4\pm0.2$\\
A119  &$335.0\pm14.9$&$0.51\pm0.01$&1.4 &$2.5\pm0.2$&$688.6\pm11.1$&$5.86\pm0.05$&5.0&$2.3\pm0.1$\\
A262  &$25.4\pm4.9  $&$0.44\pm0.01$&16.2&$1.5\pm0.3$&$93.2 \pm 2.6$&$6.27\pm0.08$&7.4&$1.2\pm0.1$\\
A401  &$227.5\pm21.1$&$0.60\pm0.02$&1.0 &$3.4\pm0.4$&$838.8\pm33.9$&$9.08\pm0.22$&4.5&$3.5\pm0.2$\\
A478  &$140.3\pm17.0$&$0.67\pm0.02$&4.9 &$3.4\pm0.5$&$478.1\pm11.3$&$9.81\pm0.16$&3.1&$3.2\pm0.2$\\
A496  &$53.6\pm9.1  $&$0.53\pm0.02$&15.0&$2.2\pm0.4$&$172.6\pm4.5 $&$7.50\pm0.09$&5.3&$1.8\pm0.1$\\
A644  &$199.0\pm24.9$&$0.69\pm0.03$&0.7 &$3.3\pm0.4$&$713.1\pm28.4$&$10.41\pm0.20$&4.7&$3.3\pm0.1$\\
A1060 &$51.9\pm3.1  $&$0.43\pm0.01$&2.3 &$1.7\pm0.1$&$178.1\pm 4.8$&$6.56\pm0.11$&3.0&$1.5\pm0.1$\\
A1651 &$160.1\pm17.3$&$0.62\pm0.02$&1.6 &$3.1\pm0.4$&$606.2\pm22.2$&$9.40\pm0.26$&2.5&$3.0\pm0.1$\\
A1795 &$108.2\pm15.5$&$0.64\pm0.02$&4.2 &$3.0\pm0.4$&$340.8\pm 6.5$&$8.92\pm0.14$&3.8&$2.6\pm0.1$\\
A2029 &$105.1\pm18.0$&$0.60\pm0.02$&2.8 &$3.5\pm0.6$&$371.2\pm 9.6$&$8.96\pm0.14$&1.8&$3.1\pm0.2$\\
A2052 &$58.6\pm10.2 $&$0.58\pm0.02$&7.7 &$2.0\pm0.4$&$174.8\pm 7.2$&$7.92\pm0.27$&2.0&$1.7\pm0.1$\\
A2063 &$84.0\pm13.2 $&$0.49\pm0.01$&7.9 &$2.0\pm0.3$&$409.6\pm14.8$&$8.40\pm0.20$&3.2&$2.1\pm0.1$\\
A2199 &$73.5\pm10.3 $&$0.58\pm0.02$&7.6 &$2.2\pm0.3$&$244.8\pm 4.7$&$8.24\pm0.08$&1.7&$2.0\pm0.1$\\
A2319 &$150.0\pm15.6$&$0.45\pm0.01$&1.1 &$3.1\pm0.3$&$538.3\pm15.5$&$7.09\pm0.15$&3.1&$3.1\pm0.1$\\
A2597 &$83.4\pm16.3 $&$0.58\pm0.03$&30  &$2.5\pm0.6$&$370.8\pm12.9$&$9.77\pm0.31$&22&$2.5\pm0.2$\\
A3266 &$288.6\pm14.4$&$0.52\pm0.01$&17.1&$3.1\pm0.2$&$773.8\pm23.4$&$6.94\pm0.09$&12.2&$3.0\pm0.1$\\
A3558 &$100.1\pm11.6$&$0.48\pm0.01$&18.2&$2.4\pm0.2$&$620.1\pm15.6$&$8.14\pm0.14$&9.6&$2.5\pm0.1$\\
A3562 &$145.2\pm11.7$&$0.51\pm0.01$&2.6 &$2.0\pm0.3$&$372.0\pm15.4$&$7.32\pm0.16$&1.6&$2.0\pm0.3$\\
A3571 &$137.8\pm11.1$&$0.54\pm0.01$&1.8 &$2.8\pm0.2$&$558.9\pm15.9$&$8.78\pm0.15$&8.4&$2.9\pm0.1$\\
A4059 &$100.8\pm14.5$&$0.59\pm0.02$&2.0 &$2.3\pm0.3$&$500.2\pm15.9$&$9.73\pm0.30$&2.0&$2.4\pm0.1$\\
MKW3  &$97.2\pm13.7 $&$0.64\pm0.02$&1.5 &$2.1\pm0.4$&$297.9\pm10.0$&$8.70\pm0.19$&6.0&$1.9\pm0.2$\\
\hline
\end{tabular}
\end{scriptsize}
\end{center}
\end{table}
}

\clearpage

\begin{figure}[htb]
\begin{minipage}{\textwidth}
\centerline{
\leavevmode\psfig{figure=./f1.ps,angle=-90,width=0.5\textwidth}
}
\end{minipage}
\end{figure}

\clearpage

\begin{figure}[htb]
\begin{minipage}{\textwidth}
\centerline{
\leavevmode\psfig{figure=./f2.ps,angle=-90,width=0.5\textwidth}
}
\end{minipage}
\end{figure}

\clearpage

\begin{figure}[htb]
\begin{minipage}{\textwidth}
\leavevmode\psfig{figure=./f3a.ps,angle=-90,width=0.5\textwidth}
\leavevmode\psfig{figure=./f3b.ps,angle=-90,width=0.5\textwidth}
\end{minipage}
\end{figure}

\clearpage

\begin{figure}[htb]
\begin{minipage}{\textwidth}
\centerline{
\leavevmode\psfig{figure=./f4.ps,angle=-90,width=0.5\textwidth}
}
\end{minipage}
\end{figure}

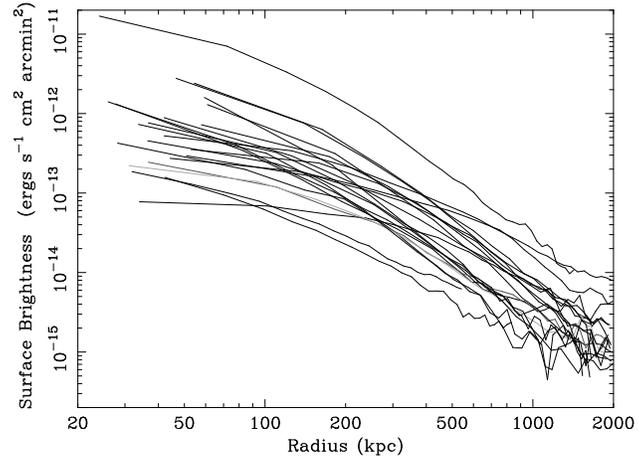
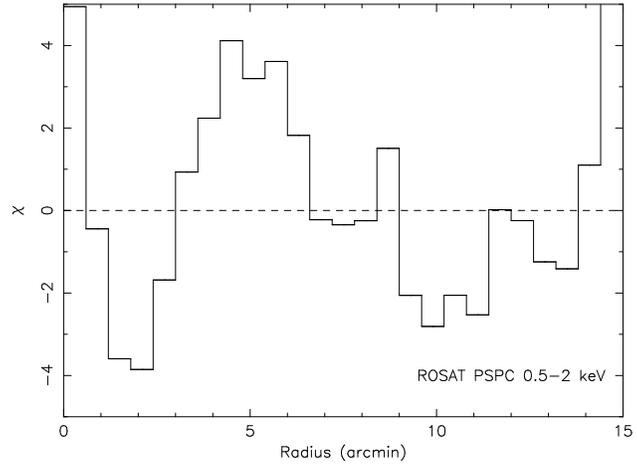
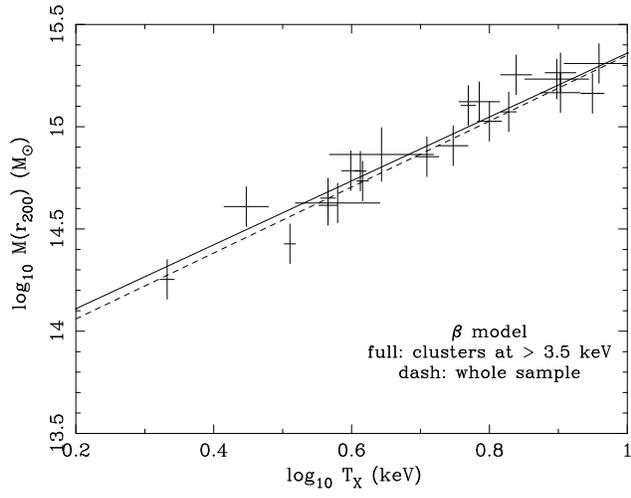
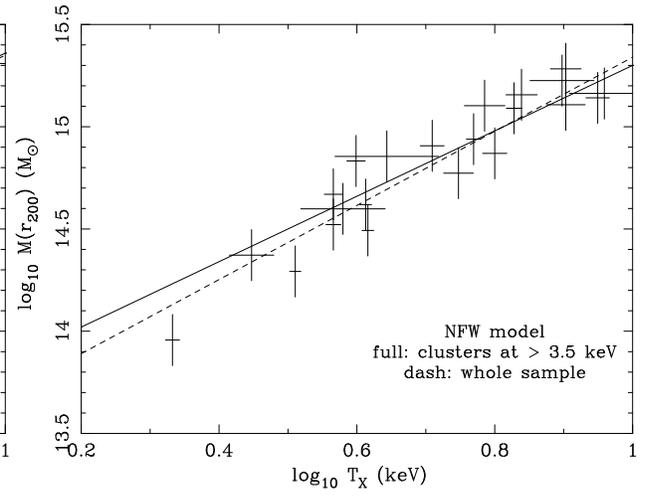
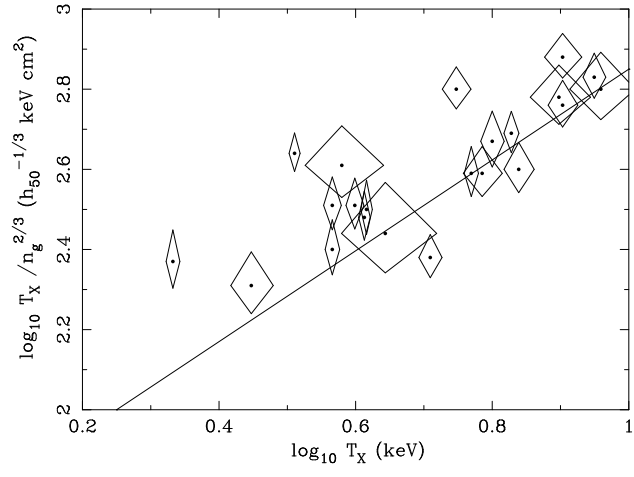
\end{document}